\documentstyle[11pt]{article}

\begin{document}

\setcounter{page}{1}

\pagestyle{plain} \vspace{1cm}
\begin{center}
\Large{\bf Tachyon field inflation in the light of BICEP2 }\\
\small \vspace{1cm} {\bf Kourosh
Nozari\footnote{knozari@umz.ac.ir}}\quad and\quad {\bf Narges
Rashidi\footnote{n.rashidi@umz.ac.ir}}\\
\vspace{0.25cm}
Department of Physics, Faculty of Basic Sciences,\\
University of Mazandaran,\\
P. O. Box 47416-95447, Babolsar, IRAN
\end{center}

\vspace{1cm}
\begin{abstract}
We study tachyon field inflation in the light of the
Planck+WMAP+BICEP2+BAO joint data. While the minimally coupled
tachyon field inflation is consistent with the Planck2013 data, it
is not confirmed by the Planck+WMAP+BICEP2+BAO dataset. However, a
nonminimally coupled tachyon field inflation is consistent with this joint dataset.\\
{\bf PACS}: 98.80.Bp,\, 98.80.Cq,\, 98.80.Es \\
{\bf Key Words}: Inflation, Tachyon Field, Observational Data
\end{abstract}
\newpage

\section{Introduction}
Inflation is a successful paradigm to solve the main problems of the
standard cosmological models such as the flatness, horizon and
relics problems. The simplest inflationary model is described by a
scalar field which slowly rolls down its
potential~\cite{Gut81,Lin82,Lyt09}. One of the scalar fields which
can be responsible for the early time inflation in the history of
the universe, is the tachyon field~\cite{Sen99,Sen02,Fei02, Noz13}
described by the Dirac-Born-Infeld (DBI) action. There are a wide
range of potentials corresponding to this field. So, we confront
with several numbers of inflationary models, which their viability
should be specified. If a model is consistent with observational
data, it can be considered as a viable inflationary model. The
recently released observational data, from Background Imaging of
Cosmic Extragalactic Polarization (BICEP2)~\cite{Ade14}, imposes new
constraints on the model's parameters. In this regard, some models
are confirmed by observation and some others are ruled out.

Two important parameters in an inflationary models are
tensor-to-scalar ratio and the scalar spectral index, which describe
the properties of the cosmological perturbations. The limit on the
tensor-to-scalar ratio from the combined WMAP9+eCMB+BAO+H$_{0}$
data~\cite{Hin13}, is $r<0.13$. Also, this combined dataset gives
the value of the scalar spectral index as $n_{s}=0.9636 \pm 0.0084$.
The joint Planck+WMAP9+BAO data~\cite{Ade13} imposes the constraint
$r<0.11$ on the tensor to scalar ratio and gives $n_{s}=0.9643 \pm
0.0059$. Now, the Planck+WMAP+BICEP2+BAO dataset ~\cite{Ade14,
Wu14}, gives new values of these parameters as
$r=0.2096^{+0.0443}_{-0.0608}$ and
$n_{s}=0.9653^{+0.0129}_{-0.0146}$. The large values of $r$,
reported by BICEP2, shows the large amplitude of gravitational wave
modes generated during inflation. Another cosmological perturbation
parameter which has attracted much interest is the spectral index of
the gravitational wave (the tensor spectral index). The
Planck+WMAP+BICEP2+BAO dataset gives
$n_{t}=1.5261^{+3.4739}_{-3.5261}$. So, this dataset shows the blue
tilt of the tensor spectral index which means that the stress-energy
tensor violates the Null Energy Condition, meaning that
$(\rho+p)<0$, so this is a challengeable result of BICEP2 data.
Also, this result doesn't satisfy the consistency relation $n_{t} =
-r/8$, predicted by the standard inflationary model. But, it is
demonstrated in~\cite{Wu14} that the spectrum of the primordial
gravitational waves have significant effect on the CMB TT spectrum
at very large scales. Actually, a large change in the positive
$n_{t}$ only induces very small change at small angular power. Since
at large scales B-mode power is absent, we can not constrain $n_{t}$
tightly with the current observational data. So, it seems that to
obtain a reliable value for the tensor spectral index, it should be
checked with future experiments.

Here we investigate the status of the tachyon field inflation in the
light of recent observational data. By exploring the
tensor-to-scalar ratio, scalar spectral index, its running and
tensor spectral index, in the background of Planck+WMAP+BICEP2+BAO
data, we find that a minimally coupled tachyon field inflation has
severe tension with observation and is disproved by recent data.
However, a nonminimally coupled tachyon field with some appropriate
potentials is consistent with recent observational data.

\section{Minimally Coupled Tachyon Field Inflation}
The action of a simple inflationary model with a tachyon field,
associated with unstable D-branes, can be written as follows
\begin{equation}
S=\int\sqrt{-g}\Bigg[\,\frac{1}{\kappa^{2}}R-V(\theta)\sqrt{1-\partial^{\mu}\theta\partial_{\mu}\theta}\,\Bigg]d^{4}x\,,
 \label{1}
\end{equation}
where $\theta$ is the tachyon field, $V(\theta)$ is its potential
and $R$ is the 4-dimensional Ricci scalar. In a FRW background, the
action (1) leads to the following Friedmann equation
\begin{equation}
\label{7}
H^{2}=\frac{\kappa^{2}}{3}\,\frac{V(\theta)}{\sqrt{1-\dot{\theta}^{2}}}\,.
\end{equation}
By variation of the action (1) with respect to the tachyon field we
obtaion the equation of motion of the tachyon field as follows
\begin{equation}
\label{8}
\frac{\ddot{\theta}}{1-\dot{\theta}^{2}}+3H\dot{\theta}+\frac{V'}{V}=0\,,
\end{equation}
where a prime marks a derivative with respect to the tachyon field
and a dot refers to derivative with respect to the cosmic time . The
energy conservation equation of the model with minimally coupled
tachyon field is given by
\begin{equation}
\label{9} \dot{\rho}_{\theta}+3H(\rho_{\theta}+p_{\theta})=0\,,
\end{equation}
where
\begin{equation}
\label{4}
\rho_{\theta}=\frac{V(\theta)}{\sqrt{1-\dot{\theta}^{2}}}\,,
\end{equation}
and
\begin{equation}
\label{5} p_{\theta}=-V(\theta)\sqrt{1-\dot{\theta}^{2}}\,.
\end{equation}

For a tachyonic inflationary model the slow roll parameters, which
are defined as $\epsilon=-\frac{\dot{H}}{H^{2}}$ and
$\eta=-\frac{1}{H}\frac{\ddot{H}}{\dot{H}}$, take the following
form~\cite{Noz13}
\begin{equation}
\epsilon=\frac{1}{2\kappa^{2}}\frac{V'^{2}}{V^{3}}\,,
\end{equation}
and
\begin{equation}
\eta=\frac{1}{\kappa^{2}}\left(\frac{V''}{V^{2}}-\frac{1}{2}\frac{V'^{2}}{V^{3}}\right)\,,
\end{equation}
Also, the number of e-folds is given by
\begin{equation}
N\simeq-\int_{\theta_{hc}}^{\theta_{f}}\kappa^{2}\frac{V^{2}}{V'}d\theta\,.
\end{equation}
where $\theta_{hc}$ denotes the value of $\theta$ at the horizon
crossing of the scales and $\theta_{f}$ is the value of $\theta$ at
the end of inflation.

To test the viability of an inflationary model, studying the
spectrum of the perturbations which are produced due to quantum
fluctuations of the fields is useful. With a perturbed FRW metric in
a longitudinal gauge as~\cite{Bar80,Muk92,Ber95}
\begin{equation}
ds^{2}=-\big(1+2\Phi\big)dt^{2}+a^{2}(t)\big(1-2\Psi\big)\delta_{i\,j}\,dx^{i}dx^{j},\label{13}
\end{equation}
the scalar spectral index, the tensor-to-scalar ratio and the
running of the scalar spectral index corresponding to a minimally
coupled tachyon model, within the slow-roll approximation
($\dot{\theta}^{2}\ll 1$ and $\ddot{\theta}\ll|3H\dot{\theta}|$),
are given as
\begin{equation}
n_{s}-1=-6\epsilon+2\eta,
\end{equation}
\begin{equation}
r=\frac{32}{25\pi}\epsilon\,,
\end{equation}
and
\begin{equation}
\alpha=\frac{dn_{s}}{d\ln k}
\end{equation}

The tensor spectral index in a minimally coupled tachyon model is
given by the following expression
\begin{equation}
\label{43} n_{t}=-2\epsilon\,.
\end{equation}

Now, we are going to test this simple model in confrontation with
observational data. To this end, we choose several potentials and
solve the integral of equation (9). Then we find the value of the
scalar field at the horizon crossing and substitute it in equations
(11)-(14). Finally we plot the behavior of $r$, $n_{s}$, $\alpha$
and $n_{t}$ in the background of Planck+WMAP+BICEP2+BAO data. The
results are shown in figure 1.

\begin{figure}[htp]
\begin{center}\includegraphics{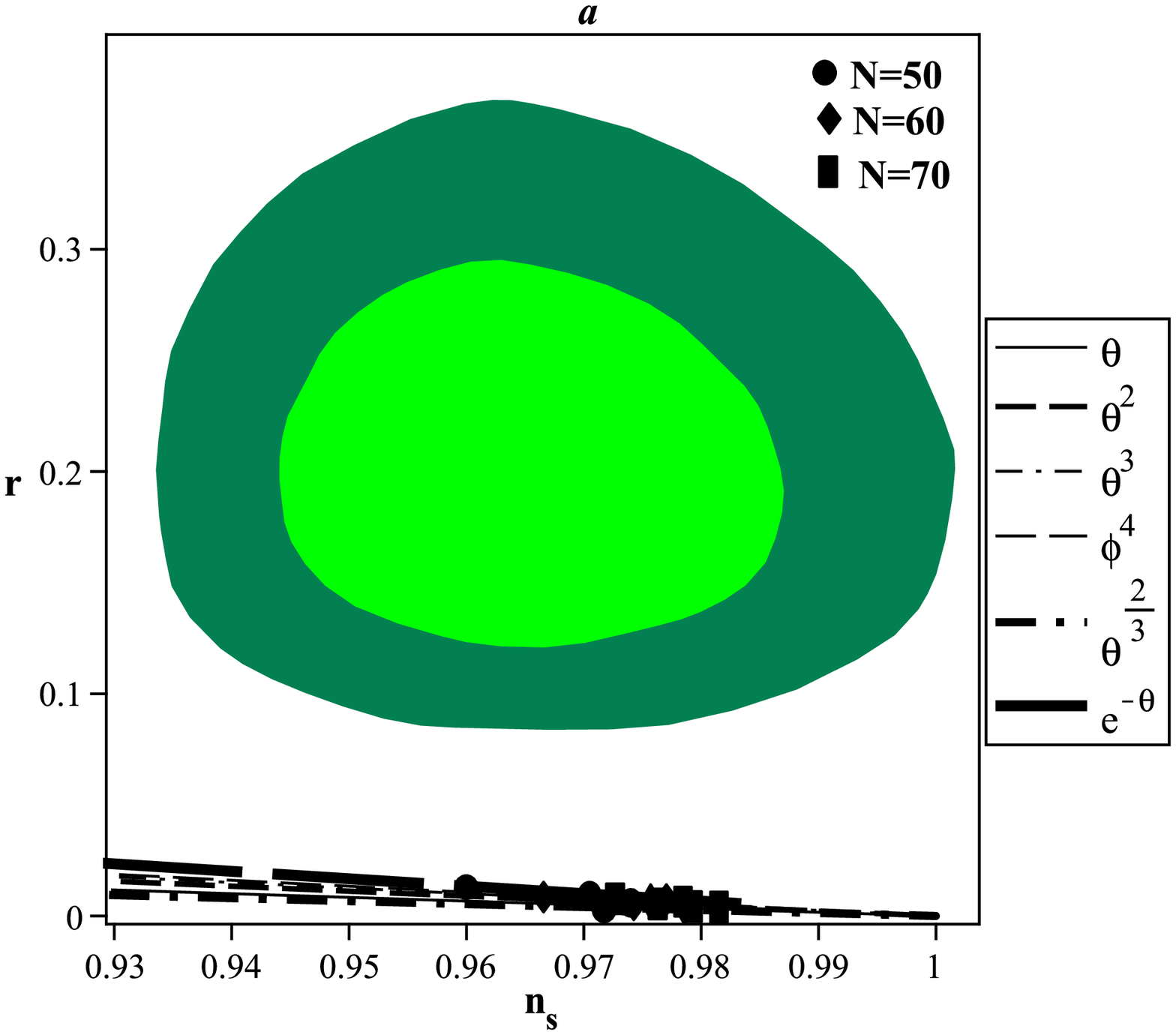}\includegraphics{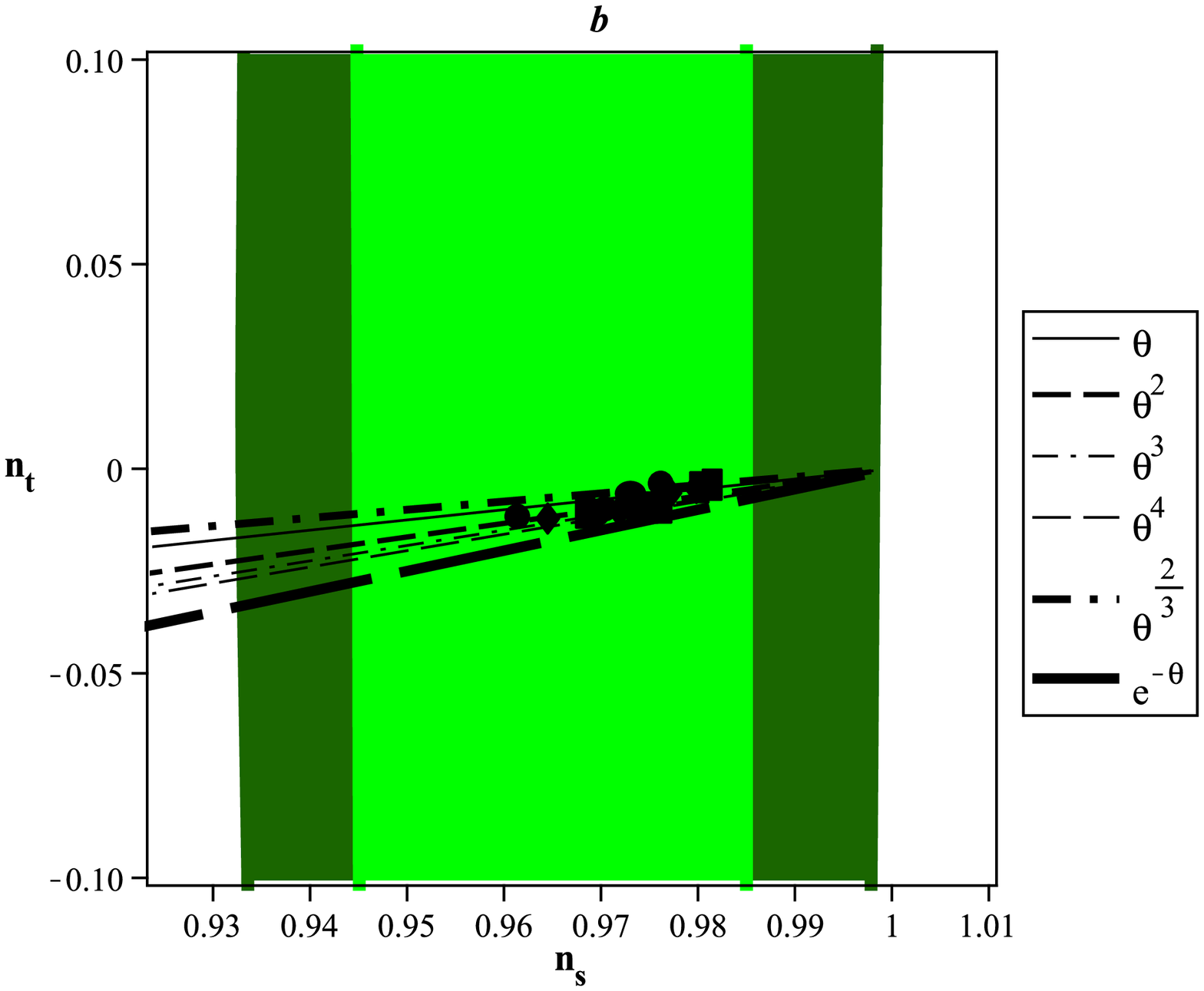} \includegraphics{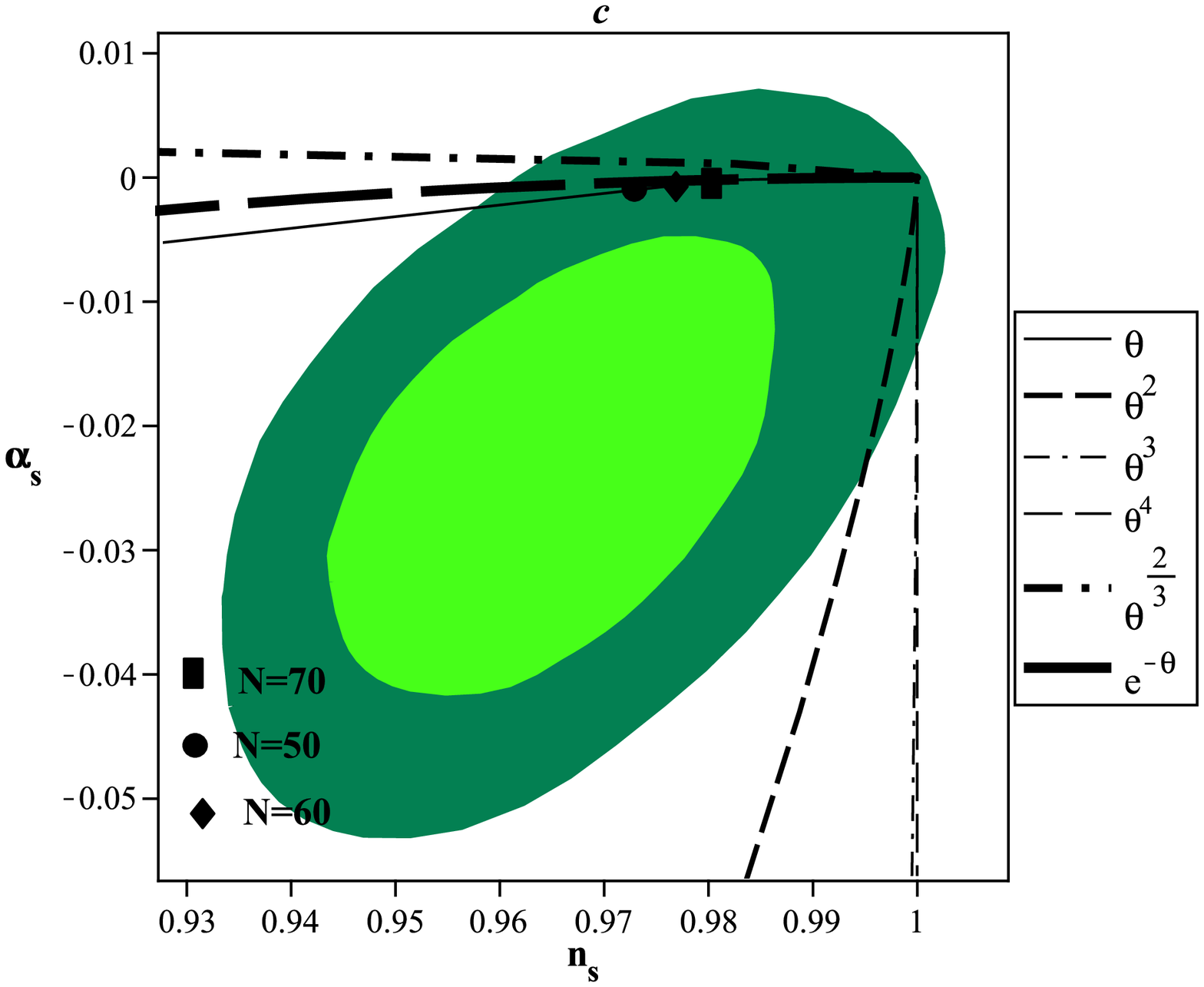}\vspace{12cm}
\end{center}
\caption{\small {Tensor-to-scalar ratio versus the scalar spectral
index (\textit{\textbf{a}}), tensor spectral index versus the scalar
spectral index (\textit{\textbf{b}}), and running of the scalar
spectral index versus the scalar spectral index
(\textit{\textbf{c}}), for a model with a minimally coupled tachyon
field and in the background of Planck+WMAP+BICEP2+BAO data.}}
\end{figure}

In panel \textit{\textbf{a}} of figure 1, we have shown the behavior
of the tensor-to-scalar ratio versus the scalar spectral index in a
model with a minimally coupled tachyon filed and with several
potentials. As the figure shows, a tachyon field model with
potentials such as a linear potential $V\sim\theta$~\cite{Mc10}, a
quadratic potential $V\sim\theta^{2}$, a potential as
$V\sim\theta^{3}$, a quartic potential $V\sim\theta^{4}$, a
potential motivated by axion monodromy, as
$V\sim\theta^{\frac{2}{3}}$~\cite{Sil08} and an exponential
potential $V\sim e^{-\theta}$ is totally outside the
Planck+WMAP+BICEP2+BAO data. We note that a minimally coupled
tachyon field inflation with potentials $\theta^{2}$ and
$e^{-\theta}$ was supported by the Planck2013 data~\cite{Noz13}. Now
we see that a minimal tachyon model is ruled out by the
Planck+WMAP+BICEP2+BAO joint data. In panel \textit{\textbf{b}} of
figure 1 we have depicted the tensor spectral index versus the
scalar spectral index. A minimally coupled tachyon field model
predicts a red tilt tensor spectral index, but the value of the
tensor spectral index is small. The evolution of the running of the
scalar spectral index versus the scalar spectral index is shown in
panel \textit{\textbf{c}} of figure 1. Once again, we see that a
minimally coupled tachyon field inflation with mentioned potentials
is not supported by this joint dataset. It should be noticed that,
with an exponential potential the running of the scalar spectral
index is far from the observational data and is not shown in the
figure. A minimally coupled tachyon model with an intermediate
potential ($V=b\,\theta^{-\beta}\,$ with $\beta=\frac{4l-4}{l-2}$,
where $l$ is defined by $a=a_{0}\exp(\vartheta t^{l})$,
$0<l<1$)~\cite{Noz13} is also quite outside the observational region
and is not shown in figure 1.

\section{Nonminimally Coupled Tachyon Field}

In the presence of a nonminimally coupled tachyon field the action
of the model is expressed as follows
\begin{equation}
S=\int\sqrt{-g}\Bigg[\frac{1}{\kappa^{2}}R-\frac{1}{2}f(\theta)R-V(\theta)\sqrt{1-\partial^{\mu}\theta\partial_{\mu}\theta}\Bigg]d^{4}x\,,
\end{equation}
where $f$ is the nonminimal coupling function defined as
$f=\xi\theta^{2}$. By using the FRW line element, the Friedmann
equation of this model is given by
\begin{equation}
H^{2}=\frac{\kappa^{2}}{3-3\kappa^{2}f}\Bigg[\frac{V(\theta)}{\sqrt{1-\dot{\theta}^{2}}}+3f'H\dot{\theta}\Bigg]\,.
\end{equation}

The equation of motion of the tachyon field obtained by varying the
action (15) with respect to the field is
\begin{equation}
\frac{\ddot{\theta}}{1-\dot{\theta}^{2}}+3H\dot{\theta}+\frac{V'}{V}+\frac{\sqrt{1-\dot{\theta}^{2}}}{2V}f'R=0\,.
\end{equation}

The presence of the nonminimal coupling changes the energy
conservation equation of the model as
\begin{equation}
\dot{\rho}_{\theta}+3H(\rho_{\theta}+p_{\theta})=-3f'H^{2}\dot{\theta}\,,
\end{equation}
where
\begin{equation}
\rho_{\theta}=\frac{V(\theta)}{\sqrt{1-\dot{\theta}^{2}}}+3f'H\dot{\theta}\,,
\end{equation}
and
\begin{equation}
p_{\theta}=-V(\theta)\sqrt{1-\dot{\theta}^{2}}-f'\ddot{\theta}-2f'H\dot{\theta}-2f''\dot{\theta}^{2}\,.
\end{equation}

The number of e-folds in this setup is given by~\cite{Noz13}
\begin{equation}
N\simeq\int_{\theta_{hc}}^{\theta_{f}}\Bigg(\frac{3V'}{V}\Bigg)\Bigg(\frac{\kappa^{2}\Big[\frac{2VV'}{3}-
\frac{f'^{2}V'R}{3V}-\frac{f'V'^{2}}{3V}\Big]}{\big(1-\kappa^{2}f\big)\big(-f'R-2V'\big)}\Bigg)d\theta\,.
\end{equation}

Now we investigate the status of a nonminimally coupled tachyon
field inflation in the light of the Planck+WMAP+BICEP2+BAO joint
data. In the presence of the nonminimal coupling and with the
perturbed FRW metric (10), the scalar spectral index of the model is
given by the following expression~\cite{Noz13}
\begin{equation}
n_{s}-1=\Bigg[\frac{\kappa^{2}\Big(f'R+2VV'\Big)}{3H^{2}V\Big(1-\kappa^{2}f\Big)}
-\frac{2V'f'R-2f''R-4V''+\frac{4V'^{2}}{V}}{-f'R-2V'}\Bigg]
\Bigg[\frac{\Big(1-\kappa^{2}f\Big)\Big(-f'R-2V'\Big)}{2V^{2}-f'^{2}R-2f'V'}\Bigg]\,.
\end{equation}
The running of the scalar spectral index is obtained by equation
(13). The tensor-to-scalar ratio is given by the following
expression~\cite{Noz13}
$$
r\equiv\frac{A_{T}^{2}}{A_{s}^{2}}\simeq\frac{8\pi}{25k^{3}}\Bigg[\frac{V^{3}-f'^{2}RV-f'V'V}{V'^{2}{\cal{C}}}\Bigg]
\hspace{9cm}\nonumber\\ \times
$$
\begin{equation}
\exp
\Bigg[-2\int\Bigg(\frac{\kappa^{2}\Big(f'R+2VV'\Big)}{6H^{2}V\Big(1-\kappa^{2}f\Big)}
-\frac{V'f'R-f''R-2V''+\frac{2V'^{2}}{V}}{-f'R-2V'}-\frac{V''}{V'}+\frac{V'}{V}\Bigg)d\theta\Bigg]
\end{equation}
In this setup the tensor spectral index takes the following form
$$
n_{t}=\Bigg(\frac{-f'R-2V'}{6H^{2}V}\Bigg)\Bigg(\frac{\kappa^{2}}{9\big(1-\kappa^{2}f\big)H^{3}}\Bigg)
\Bigg(-\frac{f'^{2}R}{2V}-\frac{f'V'}{2V}\Bigg)
$$
\begin{equation}
\times\Bigg(V'-\frac{f'f''R}{V}+\frac{f'^{2}RV'}{2V^{2}}
-\frac{f''V'}{V}+\frac{f'V''}{V}+\frac{f'V'^{2}}{V^{2}}+3\kappa^{2}H^{2}\Bigg)\,.\label{3:44}
\end{equation}

Similar to the minimally coupled tachyon field case, we consider
some potentials, solve the integral of equation (21) and by using
the equations (13) and (22)-(24), we study the behavior of $r$,
$n_{s}$, $\alpha_{s}$ and $n_{t}$ in the background of
Planck+WMAP+BICEP2+BAO data. The results are shown in the following
figures. In figure 2 we have plotted the behavior of the
tensor-to-scalar ratio versus the scalar spectral index with several
potentials and in the background of Planck+WMAP+BICEP2+BAO data.
Figure 3 shows the behavior of the tensor spectral index versus the
scalar spectral index with the same potentials. In figure 4 we see
the behavior of the running of the scalar spectral index versus the
scalar spectral index with the same potentials and in the background
of the mentioned dataset. Our numerical analysis shows that a
nonminimally coupled tachyon field inflation is observationally
viable for some sorts of potential and in some specific ranges of
the nonminimal coupling parameter.

\begin{figure}[htp]
\begin{center}\includegraphics{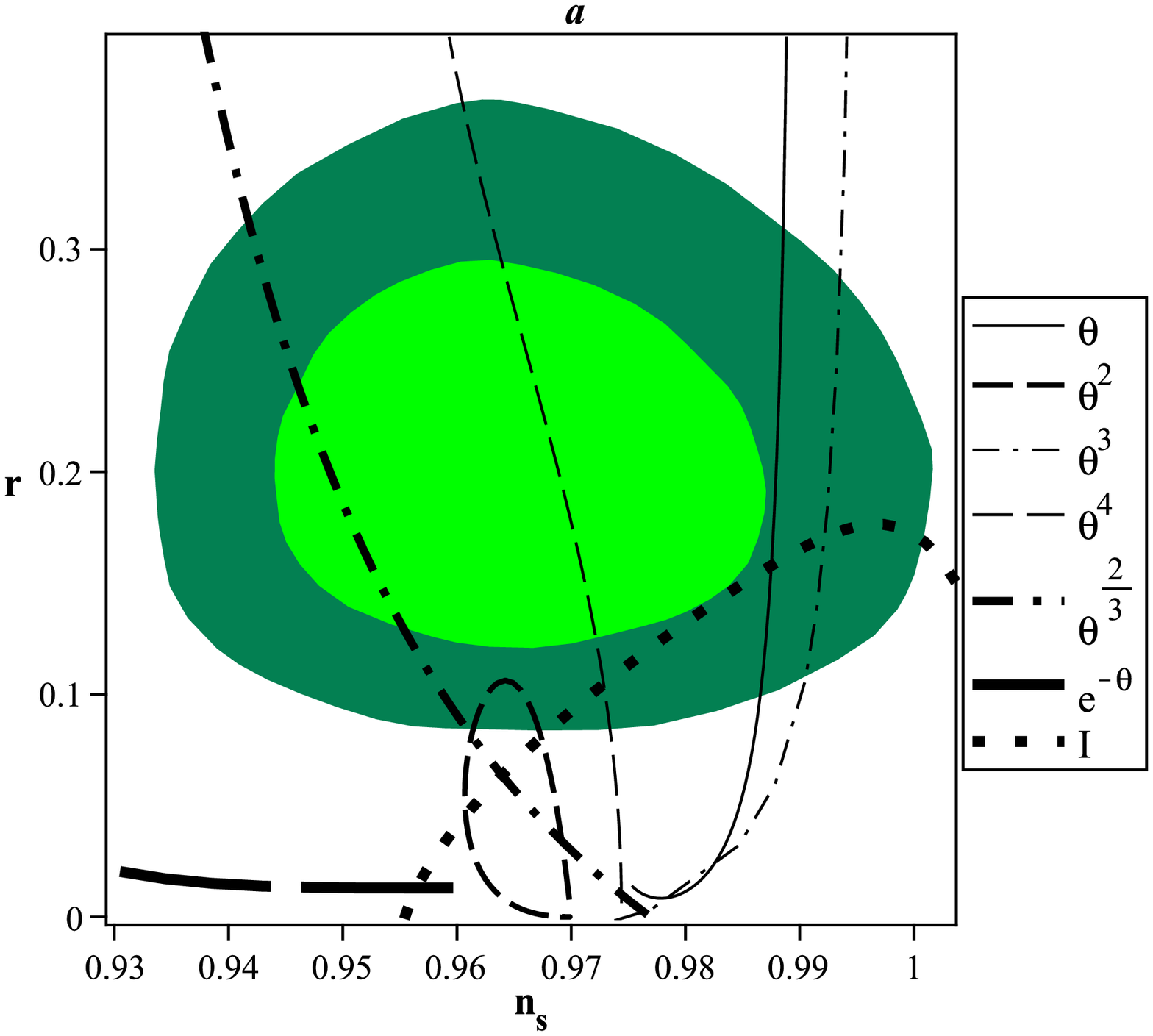}\includegraphics{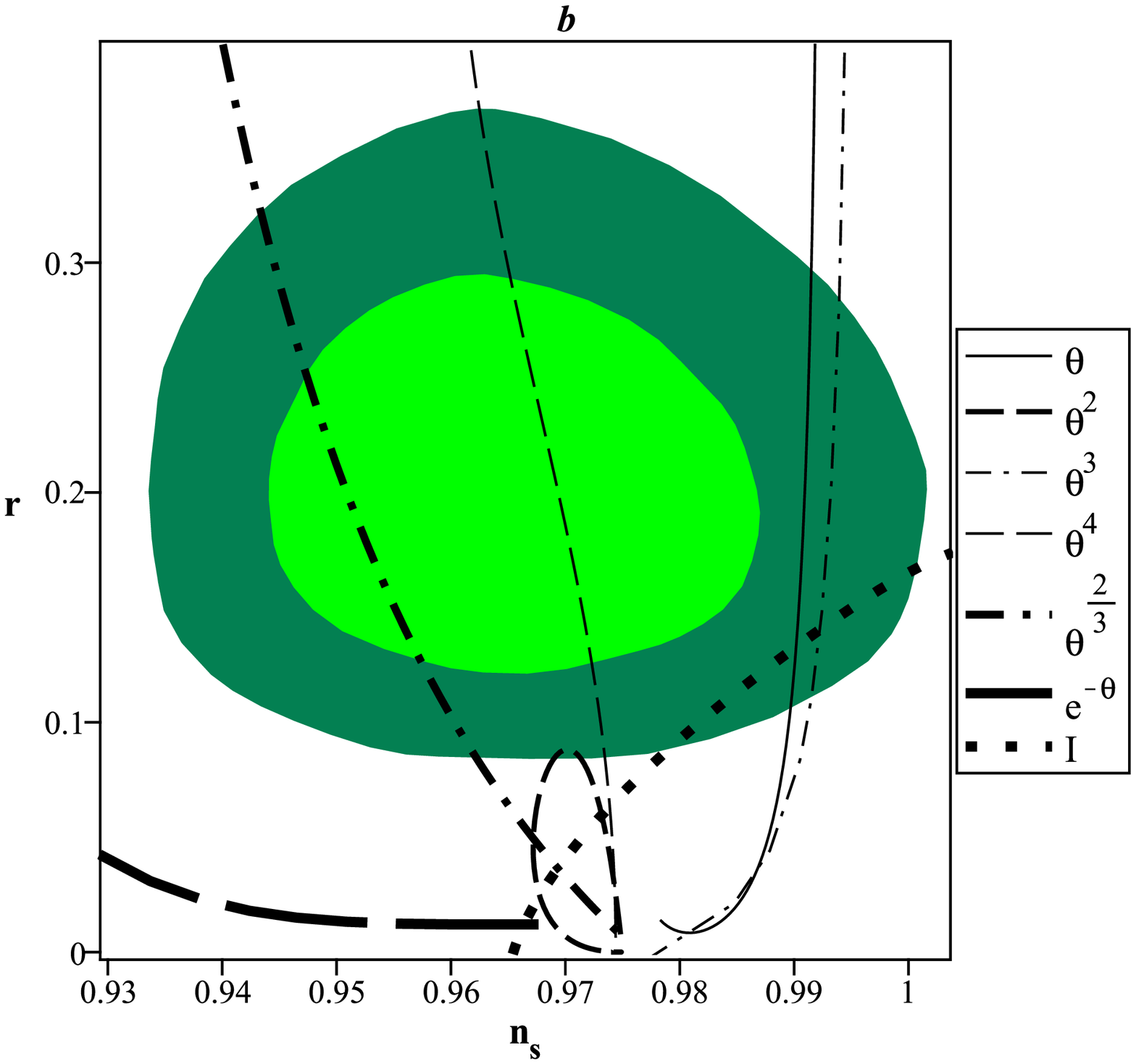} \vspace{6cm}
\end{center}
\caption{\small {Tensor-to-scalar ratio versus the scalar spectral
index for a model with a nonminimally coupled tachyon field and in
the background of Planck+WMAP+BICEP2+BAO joint data. The panels have
been plotted for $N=50$ (\textit{\textbf{a}}) and $N=60$
(\textit{\textbf{b}}). Note that, parameter $I$ in the figure refers
to \textit{Intermediate Potential}.}}
\end{figure}

\begin{figure}[htp]
\begin{center}\includegraphics{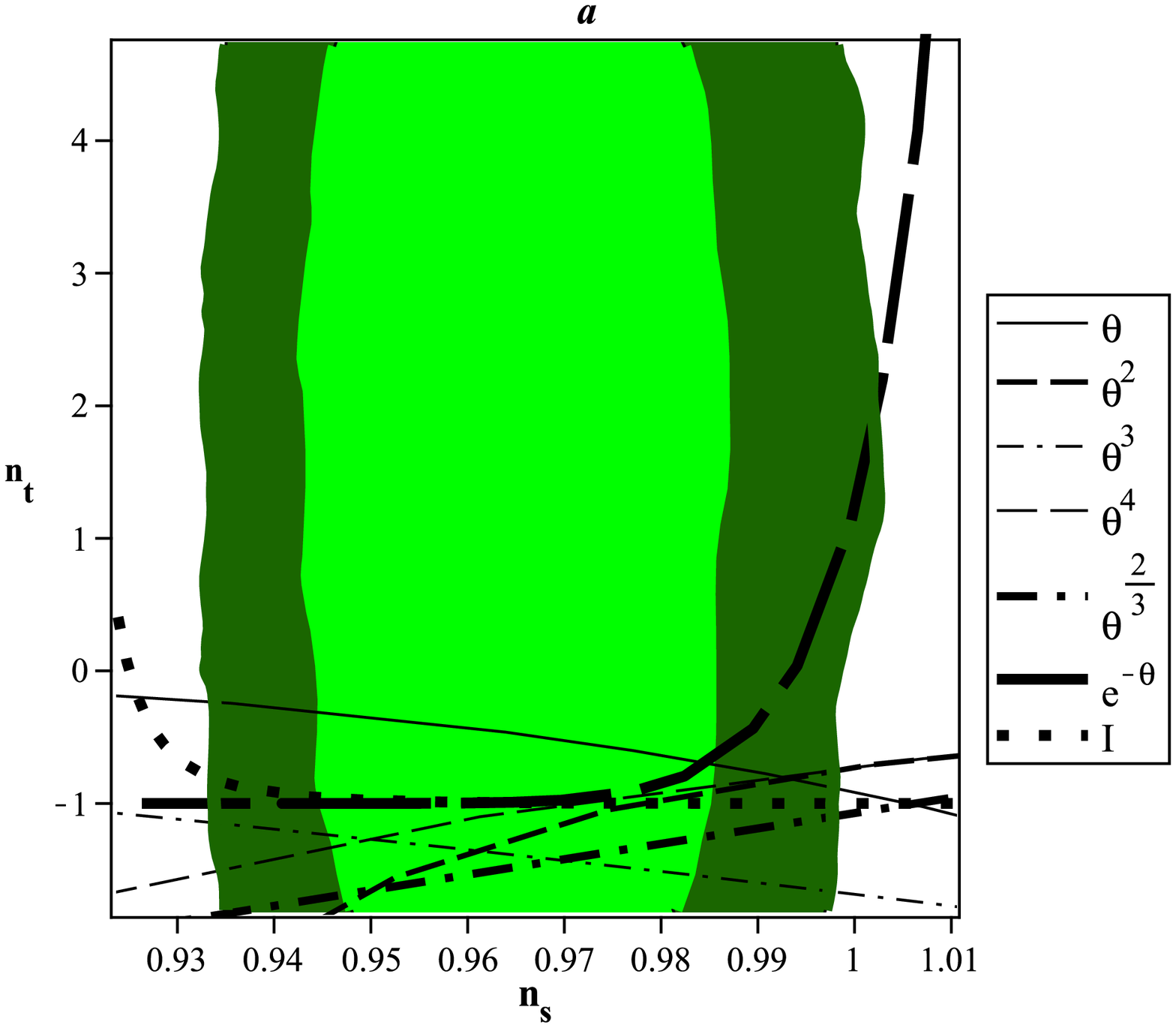}\includegraphics{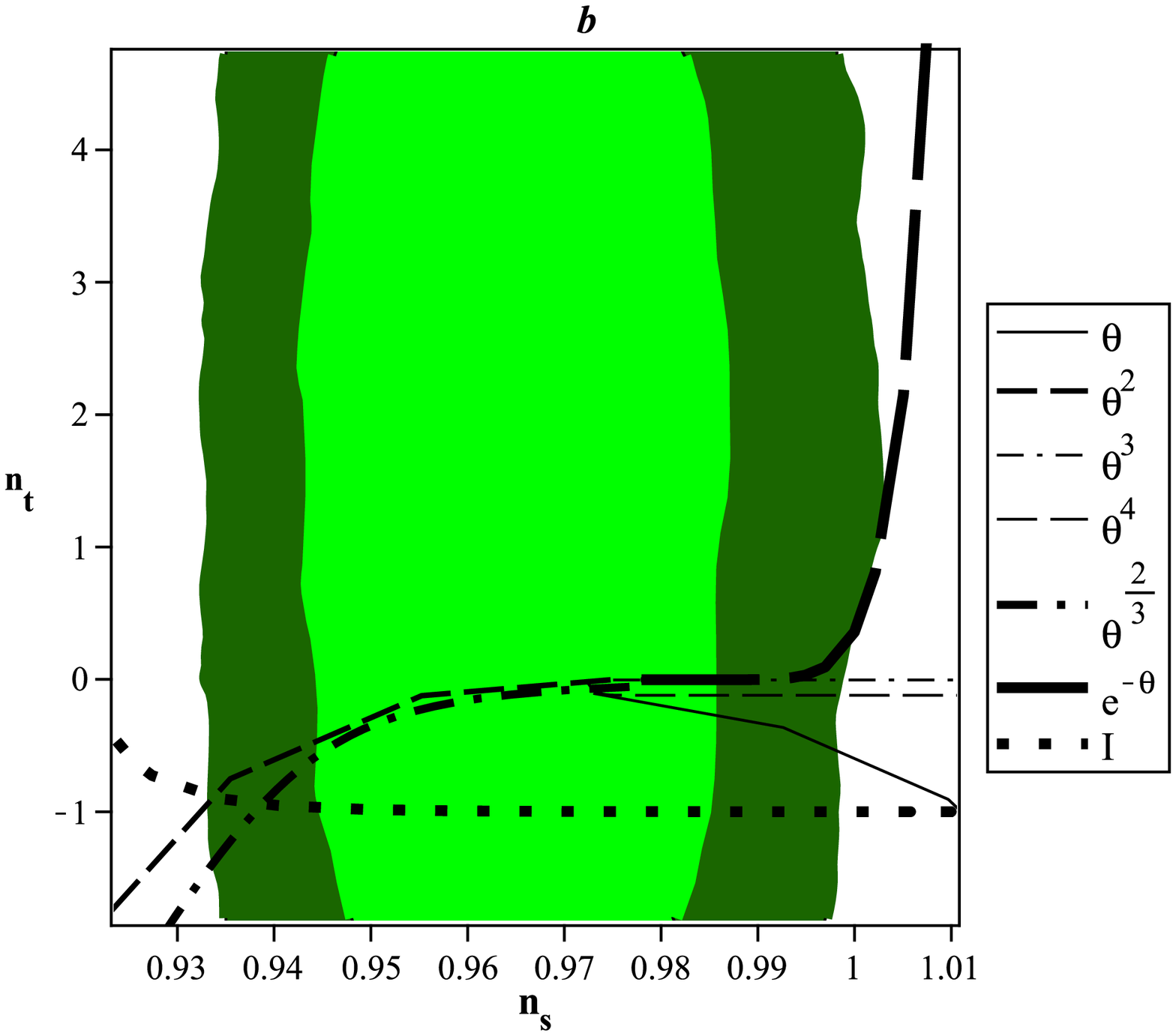} \vspace{5.5cm}
\end{center}
\caption{\small {The tensor spectral index versus the scalar
spectral index for a model with a nonminimally coupled tachyon field
and in the background of Planck+WMAP+BICEP2+BAO joint data. The
panels have been plotted for $N=50$ (\textit{\textbf{a}}) and $N=60$
(\textit{\textbf{b}}).}}
\end{figure}

\begin{figure}[htp]
\begin{center}\includegraphics{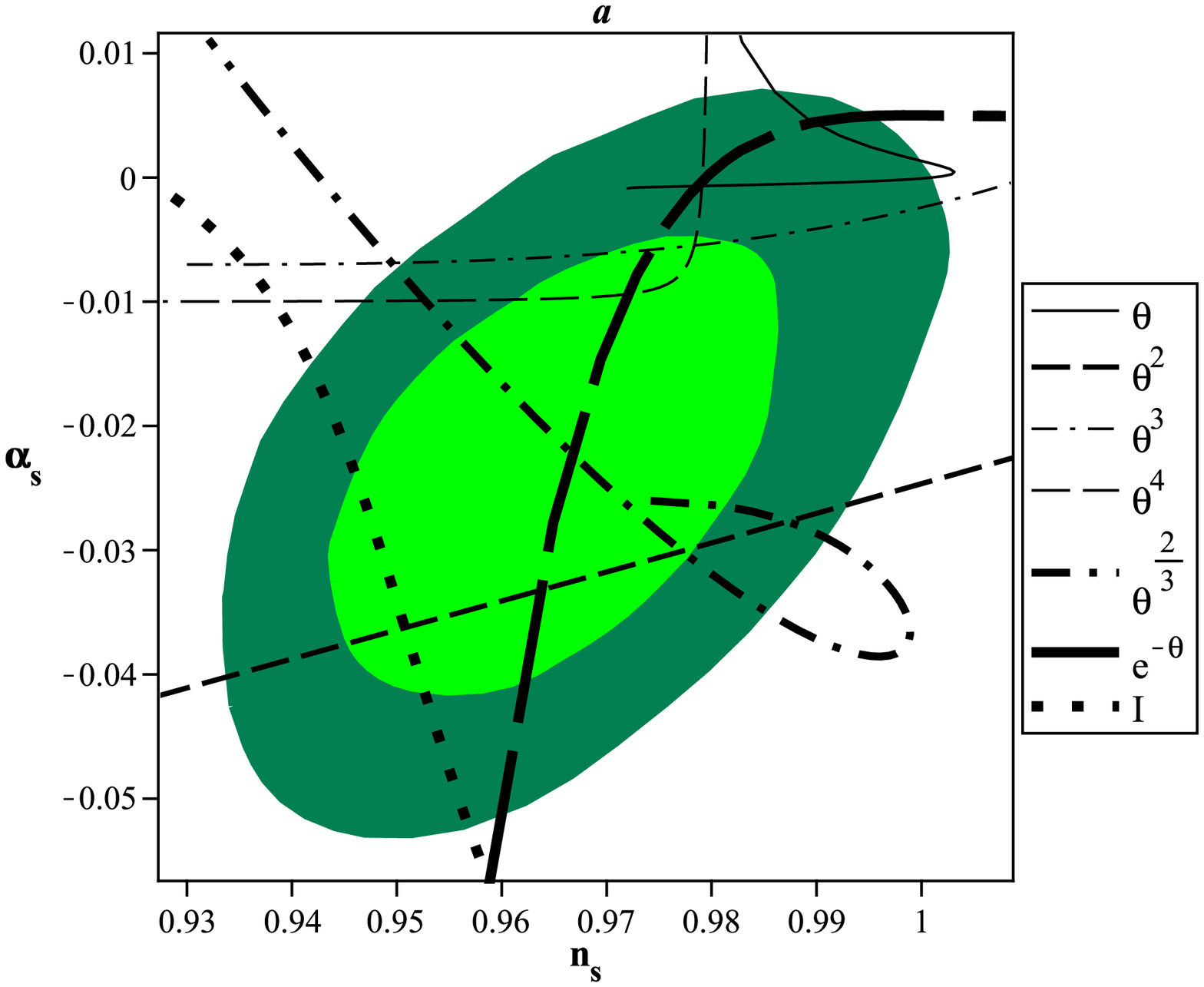}\includegraphics{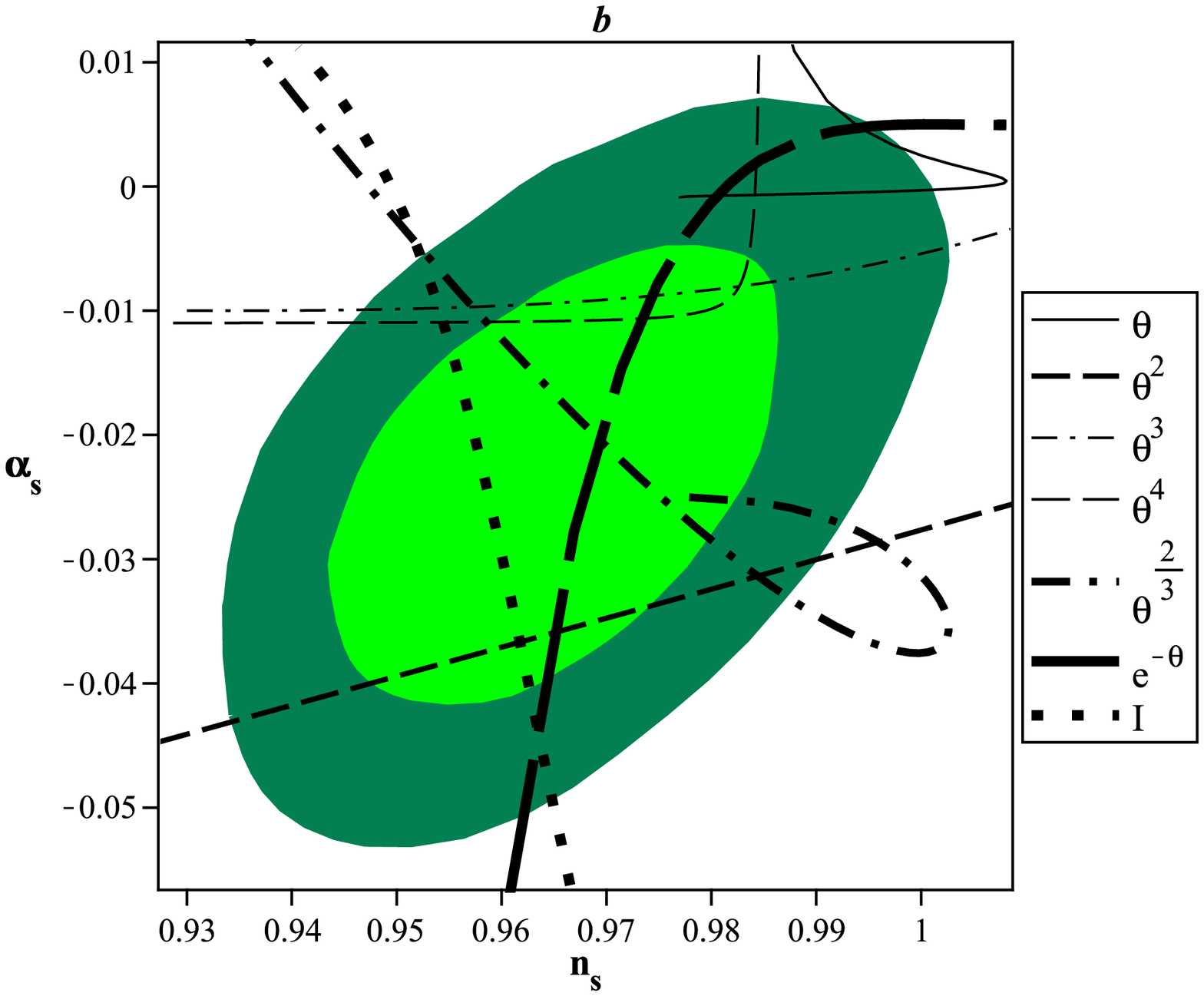} \vspace{5.5cm}
\end{center}
\caption{\small {The running of the scalar spectral index versus the
scalar spectral index for a model with a nonminimally coupled
tachyon field and in the background of Planck+WMAP+BICEP2+BAO joint
data. The panels have been plotted for $N=50$ (\textit{\textbf{a}})
and $N=60$ (\textit{\textbf{b}}).}}
\end{figure}

With a linear potential, the nonminimally coupled tachyonic model is
consistent with Planck+ WMAP+BICEP2+BAO data if $0.012<\xi<0.078$
(for $N=50$) and $0.0135\leq\xi\leq 0.0794$ (for $N=60$). A
nonminimally coupled tachyonic model with a quadratic potential is
consistent with recent observational data if $0.081<\xi<0.0883$ (for
$N=50$) and $0.0853\leq\xi\leq 0.087$ (for $N=60$). With
$V\sim\theta^{3}$, a nonminimally coupled tachyonic model is
cosmologically viable if $0.0133<\xi<0.063$ (for $N=50$) and
$0.0141\leq\xi\leq 0.0667$ (if $N=60$). With a quartic potential,
the Planck+WMAP+BICEP2+BAO data imposes the constraint
$0.0101<\xi<0.11$ (for $N=50$) and $0.0106<\xi<0.124$ (for $N=60$)
on the model. A nonminimally coupled tachyon model with a potential
as $V\sim\theta^{\frac{2}{3}}$ is observationally viable if
$0.0122<\xi<0.093$ (for $N=50$) and $0.0130\leq\xi\leq 0.0941$ (for
$N=60$). Another potential which we consider here, is an
intermediate potential. This potential with a nonminimally coupled
tachyon field can be obtained as
\begin{equation}
V=3l^{2}\vartheta^{2}\left( {\frac{b}{{\theta}^{2}}}
\right)^{\beta}\left(1
-{\kappa}^{2}\xi\theta^{2}\right)\kappa^{-2}-{\frac{\kappa^{2
}\xi^{2}{\theta}^{2+2\beta}}{1-{\kappa}^{2}\xi\theta^{2}}}\hspace{0.1cm}\nonumber\\+\xi\theta
\vartheta l{b}^{\beta}\sqrt{{\frac
{\kappa^{4}\xi^{2}\theta^{2+2\beta}}{\vartheta^{2}{
l}^{2}{b}^{\beta}\left(1-{\kappa}^{2}\xi\theta^{2} \right) ^{2}}}+6{
\frac {l-1}{\vartheta l}}}\,,
\end{equation}
where by definition
\begin{equation}
b=\frac {3 \left( -2+f \right) ^{2}\vartheta
l}{8\left(1-l\right)}\,,
\end{equation}
and
\begin{equation}
\beta=\frac {2l-2}{-2+l}\,.
\end{equation}
With an intermediate potential, there are two constrains on the
model; constrains on $\xi$ and $l$. A nonminimally coupled
intermediate tachyon field inflation is observationally viable in
some ranges of $\xi$ (as shown in the table 1), if $l>0.903$. The
constraints on the nonminimal coupling parameter are summarized in
table 1. There are two points that should be noticed here. Firstly,
with an exponential potential, a nonminimally coupled tachyon field
model is still unconfirmed by the Planck+WMAP+BICEP2+BAO data.
Secondly, as figure 3 shows, a nonminimally coupled tachyon model
with potentials $\theta$, $\theta^{2}$, $\theta^{3}$, $\theta^{4}$,
$\theta^{\frac{2}{3}}$ and also the intermediate potential predicts
a red tilt of the tensor spectral index. Only with an exponential
potential the nonminimally coupled tachyonic model predicts a blue
tilt of the tensor spectral index for $\xi<0.071$.

In summary, a minimally coupled tachyon field inflation is not
supported by the joint data of Planck+WMAP+BICEP2+BAO since these
models predict small values of the tensor-to-scalar ratio in
comparison with the recent data. However, the corresponding
nonminimal model with potentials $\theta$, $\theta^{2}$,
$\theta^{3}$, $\theta^{4}$, $\theta^{\frac{2}{3}}$ and also the
intermediate potential is supported well by the mentioned joint
dataset. Our numerical analysis sets the constraint
$0.0101<\xi<0.124$ on the nonminimal coupling parameter.

\begin{table*}
\begin{center}
\caption{The ranges of $\xi$ for which a nonminimally coupled
tachyon inflation is consistent with the Planck+WMAP+BICEP2+BAO
joint data.}
\begin{tabular}{ccccc}
\\\hline$V$&&$N=50$&&$N=60$\\ \hline
$\theta$&& $0.012<\xi<0.078$ &&$0.0135\leq\xi\leq 0.0794$\\
$\theta^{2}$&& $0.081<\xi<0.0883$ &&$0.0853\leq\xi\leq 0.087$\\
$\theta^{3}$&& $0.0133<\xi<0.063$ &&$0.0141\leq\xi\leq 0.0667$\\
$\theta^{4}$&& $0.0101<\xi<0.11$ &&$0.0106<\xi<0.124$\\
$\theta^{\frac{2}{3}}$&& $0.0122<\xi<0.093$ &&$0.0130\leq\xi\leq 0.0941$\\
$e^{-\theta}$&& --- &&---\\
$I$&& $0.041<\xi<0.093$ &&$0.046\leq\xi\leq 0.084$\\
\hline
\end{tabular}
\end{center}
\end{table*}

\end{document}